\documentclass[aps,prl,twocolumn,superscriptaddress]{revtex4}

\usepackage[colorlinks=true,urlcolor=blue,citecolor=blue,linkcolor=blue]{hyperref}
\usepackage{graphicx}
\usepackage{latexsym}
\usepackage{stmaryrd}
\usepackage{amsmath}
\usepackage{amssymb}
\usepackage{epstopdf}
\usepackage{slashed}
\usepackage{mathrsfs}
\usepackage{bm}
\usepackage{verbatim}
\usepackage{hyperref}
\usepackage[normalem]{ulem}
\DeclareMathOperator{\diag}{diag}
\DeclareMathOperator*{\argmax}{arg\,max}  

\begin{document}
\makeatletter
\newcommand{\rmnum}[1]{\romannumeral #1}
\newcommand{\Rmnum}[1]{\expandafter\@slowromancap\romannumeral #1@}
\newcommand{\B}[1]{{\textcolor{blue}{#1}}}
\newcommand{\R}[1]{{\textcolor{red}{#1}}}
\makeatother

\title{Dynamical transition in non-Hermitian Chern insulator}
\author{Zhi-Qiang Zhang }
\affiliation{Interdisciplinary Center for Theoretical Physics and Information Sciences (ICTPIS), Fudan University, Shanghai 200433, China}
\affiliation{School of Physical Science and Technology, Soochow University, Suzhou 215006, China}
\author{Yuan-Hang Ren }
\affiliation{School of Physical Science and Technology, Soochow University, Suzhou 215006, China}
\author{Ming Lu }\email{luming@baqis.ac.cn}
\affiliation{Beijing Academy of Quantum Information Sciences, Beijing 100193, China}
\author{Hua Jiang}
\affiliation{Interdisciplinary Center for Theoretical Physics and Information Sciences (ICTPIS), Fudan University, Shanghai 200433, China}
\date{\today}
	
\begin{abstract}
We unveil a peculiar dynamical transition for the propagation of wave packets in non-Hermitian Chern insulators, where the evolution of the topological wave packets at the edge is not solely determined by the topological Chern number. Unlike the Hermitian Chern insulator, where a wave packet initiated at the edge propagates along the system boundary, here it may instead penetrate into the bulk. This behavior is attributed to the competition between the localization induced by conventional topology and the non-Hermitian skin effect. Specifically, when the former dominant, the wave packet will evolve along the boundary; otherwise, it will spread into the bulk. These features demonstrate that while the generalized-Brillouin zone framework reliably predicts topological phase transitions, it does not on its own specify the dynamical transitions.
\end{abstract}
	
\maketitle

\section{I. Introduction}
 Over the past few years, the study on the topological property has been expanded from Hermitian to non-Hermitian systems \cite{zongshu1,zongshu2,zongshu3,NH1,NH2,NH3,NH4,NH5,NH7,NH8,NH9}. As one of the most significant features of non-Hermitian systems, the non-Hermitian skin effects (NHSEs) have attracted great interest in both theory and experiments \cite{ref6,ref7,ref8,ref9,ref10,ref11,ref12,ref13,ref14,ref15,ref16,ref17,ref18,ref19,ref20,ref21,ref22,ref23,ref24,ref25,ref26,NHHOTI_ex,XiongY,WER, jjLiu}.
The NHSE is characterized by the unexpected localization of eigenstates, leading to the breakdown of the conventional bulk-boundary correspondence in non-Hermitian topological systems \cite{NHSE1,NHSE2,NHSE3,NHSE4,MGBZOS,MGBZOS2}.
Notably, this issue has been resolved by the generalized-Brillouin zone (GBZ) theory \cite{NHSE1,NHSE2,NHSE3}, which appropriately redefines topological invariants on the GBZ.

Based on the GBZ theory, previous studies on non-Hermitian topological systems have primarily focused on their static properties \cite{NHSE1,NHSE2,NHSE3,NHSE4}. Dynamical evolutions of wave packets \cite{dynamicNH,dynamic0,dynamic1,dynamic2,dynamic3,dynamic4,dynamic5,dynamic6,dynamic7}, despite being a key inspiration for Hermitian topological systems, have rarely been explored in non-Hermitian topological systems. In Hermitian systems, a robust correspondence between dynamical and static characteristics is guaranteed, where the dynamical evolutions of topological edge states are directly determined by their static topological invariants. For example, in a Chern insulator, the wave packet showing a unique chiral motion along the sample edge as long as the topological invariant is nontrivial \cite{CIexperiments1,CIexperiments2,CIexperiments3}. However, due to the unique properties of non-Hermitian systems, the correspondence between the static properties and dynamical responses might be fundamentally different from that of Hermitian cases, yet remains understudying.

\begin{figure}[t]
\centering
\includegraphics[width=0.49\textwidth]{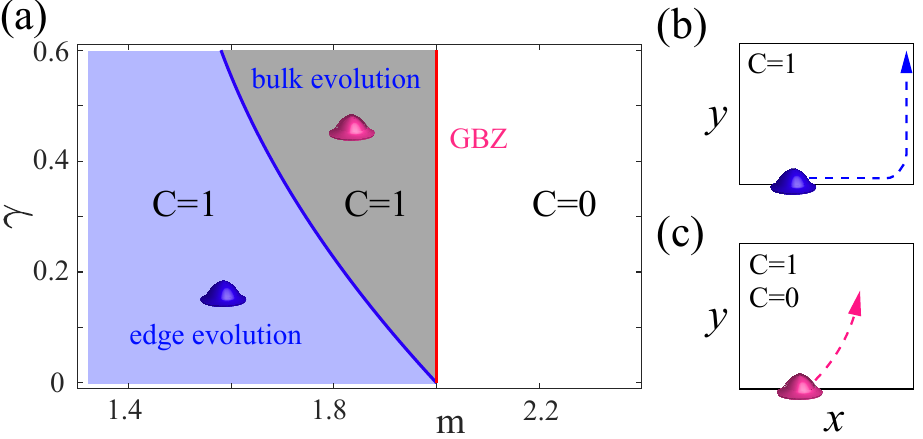}
	\caption{(a) The phase diagram. The topologically trivial white region with Chern number $C=0$ is separated with the topologically nontrivial gray and blue regions, by the red curve predicted by the GBZ theory. The solid blue curve delineates the distinct dynamical regimes marked in (b) and (c), where the wave packets evolve along the boundary in (b) , and spread into the bulk in (c).
}
	\label{f1}
\end{figure}

In this work, we unveil a distinctive static-dynamical correspondence in non-Hermitian Chern insulators: in contrast to its Hermitian counterparts, the topological invariant alone cannot determine the wave packet evolution at the boundary. In typical pseudo-Hermitian cases with negligible imaginary spectra \cite{PHsystem}, the dynamical evolution of wave packets manifests two distinct schemes, arising from the competition between the topologically protected intrinsic localization of edge states and the NHSE.
When intrinsic localization dominates, the edge states display conventional localization properties, and the wave packet propagates along the edge of the sample, as illustrated in Fig.~\ref{f1}(b). In contrast, when the NHSE dominates, the localizations of edge states are similar to their bulk states. In this case, the wave packet spreads into the bulk even though a nontrivial Chern number still exits, as shown in Fig.~\ref{f1}(a) and Fig.~\ref{f1}(c). Our findings reveal a unique interplay between bulk topology and NHSE in shaping the dynamical properties of non-Hermitian Chern insulators, paving the way for understanding the emergent dynamical phenomena in non-Hermitian topological systems.

\section{II. Edge and bulk evolution in topological regime}
 We first present the evolution of the wave packet in non-Hermitian Chern insulators, with the following Hamiltonian adapted from Qi-Wu-Zhang (QWZ) model \cite{zongshu1,zongshu2,zongshu3}:

\begin{align}\label{model}
	\begin{split}
H(\textbf{k})&=\big[m-\sum_{i=x,y}t_i(\beta^+_i\cos k_i+i\beta^-_i\sin k_i)\big]\sigma_z\\
&\quad+\sum_{i=x,y}v(\beta^+_i \sin k_i-i\beta^-_i\cos k_i)\sigma_i.
	\end{split}
\end{align}
For this model, the relative non-Hermiticity strength is $\beta_i^-/\beta_i^+$, with $\beta^\pm_{i}=(e^{\gamma_i}\pm e^{-\gamma_i})/2$. According to standard GBZ theory, the magnitude of the GBZ is given by  $\beta_i=\sqrt{\frac{1+\beta_i^-/\beta_i^+}{1-\beta_i^-/\beta_i^+}}=e^{\gamma_i}$. Therefore, $\gamma_i=\ln\beta_i$ denotes the inverse of the localization length of the NHSE along the $i$-th direction.

The dynamical evolution of the wave packet in the system satisfies \cite{dynamicNH,NHSE2}
 \begin{equation}
|\psi(t)\rangle=e^{-iHt/\hbar}|\psi_{\text{in}}\rangle.
\end{equation}
Here, $H$ is the real-space Hamiltonian for $H(\textbf{k})$ under the open boundary condition with sizes $N\times N$. System parameters are set as $\gamma_x=\gamma_y=\gamma$ and $v=t_x=t_y=1$, and we also set $\hbar=1$ in the following for simplicity.

For an initial wave packet located at the bottom edge, namely $ \langle \bm{r}|\psi_{\text{in}}\rangle \propto e^{-\Delta_x(x-N/2)^2-\Delta_y(y-1)^2}$  with $\Delta_x=\Delta_y=1/40$. The dynamical evolution of the wave packet shows distinct features. Generally, for a Hermitian system with $\gamma=0$, the center of the wave packet will mainly be located at the boundary when the Chern number $C=1$ and spread into the bulk when $C=0$.
This is originated from the conventional bulk-boundary correspondence, where $C=1$ dictates the existence of the edge states, whose localization property determines the evolution of the wave packet along the boundary \cite{CIexperiments1,CIexperiments2,CIexperiments3}. In contrast, for $C=0$, the edge states are absent and $\langle \bm{r}|\psi(t)\rangle$ spreads into the bulk.

\begin{figure}[t]
\centering
\includegraphics[width=0.49\textwidth]{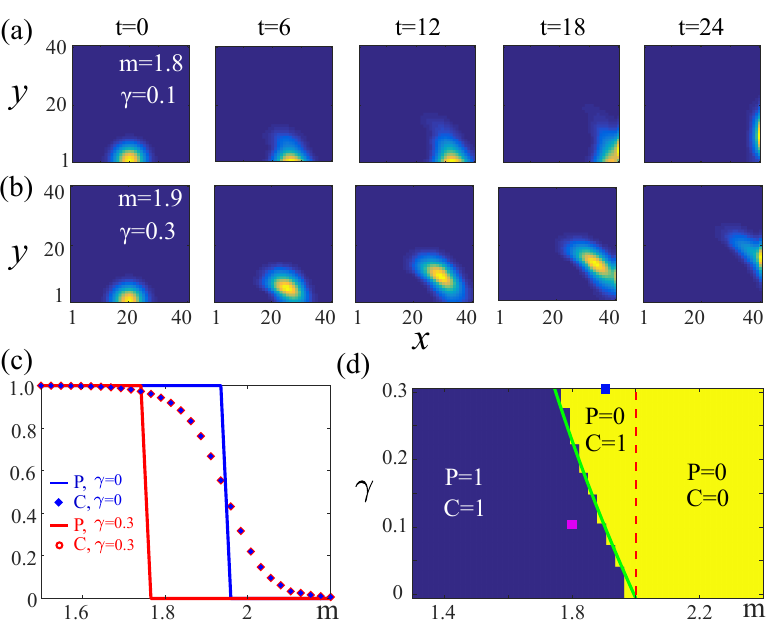}
	\caption{ (a) and (b) are the time evolutions of wave packet $|\psi(\bm{r}, t)|$ with boundary and bulk type evolutions, respectively. Note that they are both in the topological region, as marked in (d) with red and blue squares. (c) The variation of $P$ and Chern number $C$ for two representative $\gamma$'s, showing clear deviation for non-Hermitian case. (d) The plot of $P$ and $C$ versus $m$ and $\gamma$ with system size $N_x=N_y=30$. The red dashed line is the topological phase transition line based on the GBZ theory. The solid green line represents the dynamical transition line given by Eq.~\eqref{eq:dynamical_transition_line}. The parameters are $v=t_x=t_y=1$ and $\gamma_x=\gamma_y=\gamma$, with others marked in the figure accordingly.}
	\label{f4}
\end{figure}

 Nevertheless, due to the existence of the NHSE, the validity of this principle is questionable. Below, we explicitly demonstrate that the Chern number protected dynamical correspondence fails in general for the non-Hermitian Chern insulators.
 Notably, to characterize the dynamical phenomena of a system, the most intuitive approach is to directly plot the temporal evolution of wave packets. However, this method becomes inefficient for analyzing wave packet dynamics across a large parameter space.
  To clarify this issue, we introduce an index $P$ that indicates whether the wave-packet center remains at the boundaries, thereby distinguishing edge evolution from bulk evolution of the wave packet.
 For this purpose, we pay attention to the distribution of the wave packet center for different times, which is marked by the location of its maximum value
\begin{equation}
\bm{R}_c(t)=\argmax_{x,y\in [1,N]}|\psi(\bm{r},t)|^2=[x_c(t),y_c(t)],
\end{equation}
 with $\psi(\bm{r},t)=\langle \bm{r}|e^{-iHt}|\psi_{\text{in}}\rangle$. For a square sample with sizes $N\times N$, the wave packet is at the boundary if $\bm{R}_c(t)$ is within the following four cases: (i) $x_c=1$ for the left edge; (ii) $x_c=N$ for the right edge; (iii) $y_c=1$ for the bottom edge; (iv) $y_c=N$ for the top edge.
 Otherwise, the wave packet center is located in the bulk.
 More details for the calculation of $P$ is presented in Appendix II. Here, $P=1$ corresponds to the persistent edge evolution cases.
Relatively, $P=0$ corresponds to the bulk evolution, where the wave packet propagates from the boundary to the bulk. As a concrete example, the persistent edge evolution in Fig. \ref{f4}(a) and the bulk evolution in Fig. \ref{f4}(b) can be clearly identified with $P=1$ and $P=0$, respectively. It should be noted that the index $P$ defined here serves merely as a convenient descriptor of the two distinct dynamical evolutions. It is not meant to provide the predictive power, but rather to guide the subsequent discussion of the underlying mechanism. To explicitly and unbiased calculate the boundary line with only the properties of the static Hamiltonian, see Eq.~\eqref{eq:dynamical_transition_line} and related discussion in Sec.\,IV.

For comparison, the real-space Chern number $C$ is also calculated under open boundary conditions.
The Chern number is calculated by adopting the non-commutative geometry method \cite{Fsong,Liu1} with
\begin{equation}
C=-2\pi i\sum_{n= N_x/2,N_y/2}\langle n_L|P_r[-i[\widehat{x},P_r],-i[\widehat{y},P_r]] |n_R\rangle.
\label{EQ2}
\end{equation}
$|n_{L/R}\rangle$ represents the left/right eigenstate. $P_r=\sum_{E<E_F}|n_{R}\rangle \langle n_{L}|$ is the projection operator of the occupied states. $N_x$ ($\widehat{x}$) and $N_y$ ($\widehat{y}$) are the sample size (the coordinate operator) along the $x$ and $y$ directions, respectively. Following the method in Ref. \cite{Fsong,Liu1}, the Chern number is calculated under the open boundary conditions.
$C$ is adopted by considering the site Chern number at the center lattice point of the sample with $n=[N_x/2,N_y/2]$.

As shown in Figs. \ref{f4}(c), the indices $P$ and $C$ are plotted as a function of $m$ for two representative $\gamma$.
One notices that the edge dynamical evolution turns to the bulk type as $m$ increases, where $P$ varies from $1$ to $0$. The variation of $P$ is consistent with the variation of $C$ for $\gamma=0$ (blue line and dots), which illustrates the dynamical bulk-boundary correspondence in the Hermitian case. However, for $\gamma\neq0$, the variations of the two indices are inconsistent (red line and dots), showing a peculiar discrepancy between the topological and dynamical transitions in non-Hermitian systems. Due to this inconsistency, the phase diagram can be partitioned into three regions by the dynamical and topological transition lines, as shown in Fig.~\ref{f4}(d), showing the richness of the dynamical phases in non-Hermitian Chern insulators. These results demonstrate the failure of the widely adopted principle in Hermitian systems, where the dynamical properties of the edge are solely determined by the topological characteristics of the bulk. We also emphasis that for these distinctive dynamical evolution behaviors to emerge, the initial position of the wave packet should be deliberately placed at the lower or left edges of the system due to the direction of the NHSE. Otherwise, topology and non-Hermiticity will act constructively and force the wave packets stick on the edge throughout the evolution. 

\section{III. Mechanism of the dynamical transition}
In this section, we illuminate the mechanism of the peculiar dynamical transitions in non-Hermitian Chern insulators. The time evolution of the wave packet $|\psi_{\text{in}}(t)\rangle$ can be calculated by exploiting the biorthogonal expansion \cite{dynamicNH}
\begin{align}\label{eq:biorthogonal-expansion}
	\begin{split}
|\psi(t)\rangle= e^{-iHt}|\psi_{\text{in}}\rangle=\sum_ne^{-iE_n t}|n_R\rangle\langle n_L| \psi_{\text{in}}\rangle,
	\end{split}
\end{align}
where the left and right eigenvectors satisfy $H|n_R\rangle=E_n|n_R\rangle$ and $H^\dagger|n_L\rangle=E_n^*|n_L\rangle$ \cite{NH1,Liu2}, respectively. This form of expansion is valid as long as there is no exceptional degeneracy in the system. Otherwise, the effects of the exceptional points should be taken care of appropriately \cite{EP1,EP2,EP3,EP4,EP5}.

To characterize the spatial distributions of the wave packet, it is helpful to perform a similarity transformation
\begin{equation}
\mathcal{H}=S^{-1}HS,
\end{equation}
which connects the non-Hermitian Hamiltonian $H$ with its Hermitian counterpart $\mathcal{H}$. The similarity transformation ensures that $H$ and $\mathcal{H}$ possess the same spectrum and topological properties \cite{MGBZOS}.
Specifically, the eigenvectors $|\varphi_n\rangle$ of $\mathcal{H}$ satisfy $\mathcal{H}|\varphi_n\rangle=E_n|\varphi_n\rangle$, with the orthonormal condition $\langle \varphi_m|\varphi_n\rangle=\delta_{mn}$.
They are related to the left and right eigenvectors by \cite{NH1,Liu2}
 \begin{align}
	\begin{split}
  |n_R\rangle=S|\varphi_n\rangle,~~ \langle n_L|=\langle\varphi_n|S^{-1}.\label{eq:basis_transform}
  	\end{split}
\end{align}
Substituting Eq.~\eqref{eq:basis_transform} into Eq.~\eqref{eq:biorthogonal-expansion}, the time evolution of the wave packet in real space reads
\begin{equation}
\psi(\bm{r},t) =\sum_ne^{-iE_n t}\langle \bm{r}|S|\varphi_n\rangle\langle \varphi_n|S^{-1} |\psi_{\text{in}}\rangle
\end{equation}

The transformation operator $S$ for our model Eq.\eqref{model} in real space can be represented as $S=\diag [1,\cdots,[\beta_x]^{x}[\beta_y]^{y},\cdots]$, with $\beta_{x,y}=e^{\gamma_{x,y}}>1$, revealing the existence of NHSE in $H$ \cite{localization,MGBZOS}. Thus,
\begin{equation}
    \psi(\bm{r},t) =\sum_n c_n e^{-iE_n t} e^{\ln(\beta_x)x+\ln(\beta_y)y} \langle \bm{r}|\varphi_n\rangle,
\end{equation}
with the time-independent coefficient $c_n$ representing the matrix element $\langle \varphi_n|S^{-1} |\psi_{\text{in}}\rangle$ . Writing explicitly in the coordinate space,
\begin{equation}\label{eq:c_n_def}
    c_n=\langle \varphi_n|S^{-1} |\psi_{\text{in}}\rangle = \int d^2\bm{r} \varphi_n^*(\bm{r}) \tilde{\psi}_\text{in}(\bm{r}).
\end{equation}
For the studied model, $S^{-1}$ enhances wave packet localization in the lower-left direction. When the initial wave packet $\langle \bm{r}|\psi_{\text{in}}\rangle$ is lower-boundary-localized, the modified wave packet $\langle \bm{r}|\tilde{\psi}_{\text{in}}\rangle=\langle \bm{r}|S^{-1}|\psi_{\text{in}}\rangle$ exhibits enhanced lower-boundary localization.
Mathematically, $\tilde{\psi}_\text{in}(\bm{r})=e^{-\ln(\beta_x)x-\ln(\beta_y)y} \psi_\text{in}(\bm{r})$ is still a Gaussian wave packet localized at the boundary, with a modified center and amplitude by $\beta_{x,y}$ compared to the original $\psi_\text{in}(\bm{r})$.

Recall $|\varphi_n\rangle$ are the eigenstates for the Hermitian Hamiltonian $\mathcal{H}$, whose bulk and topological edge states should roughly satisfy
\begin{align}\label{wavefunction}
	\begin{split}
    \langle \bm{r}|\varphi_{n\in bulk}\rangle&\propto\frac{1}{N}e^{ik_x^{(n)}x+ik_y^{(n)}y}\\
\langle \bm{r}|\varphi_{n\in edge}\rangle&\propto\frac{1}{\sqrt{N}} e^{ik_x^{(n)}x-\kappa_yy}.
	\end{split}
\end{align}
with $N$ represents the linear dimension of the system. Here, we focus on the bottom edge states throughout
the calculations, which are extended along the $x$-direction and localized along the $y$-direction.  Similar forms apply for the edge states of other boundaries.

For the topological trivial scenario, there are no edge states, the wave packet reads
\begin{equation}\label{eq:topo_trivial}
    \psi(\bm{r},t) \propto\sum_n \frac{c_n}{N} e^{-iE_n t} e^{\ln(\beta_x)x+\ln(\beta_y)y} e^{ik_x^{(n)} x+ik_y^{(n)} y},
\end{equation}
Since $\beta_{x,y}>1$, the wave function grows exponentially with increasing $x$ or $y$. Thus, the wave packet will spread into the bulk and admit the bulk evolution.
In general, the imaginary eigenvalue $\operatorname{Im}[E_n]$ could give rise to an extra coefficient $c(E_n)=e^{\operatorname{Im}[E_n]t}$, and could significantly alter the dynamical properties of the non-Hermitian Chern insulators in the long time limit. In our deviations, we have neglected the influences
of $\operatorname{Im}[E_n]$, because the studied model Eq.~\eqref{model} possesses
the vanishingly small imaginary eigenvalues under the open boundary conditions.

In the topological regime, the topological edge states are present. As shown in Eq.~\eqref{eq:c_n_def}, the coefficient $c_n$ represents the inner product of the eigenfunction $\varphi_n(\bm{r})$ and the boundary-localized wave packet $\tilde{\psi}_\text{in}(\bm{r})$. This inner product is much larger when $\varphi_n(\bm{r})$ is a topological edge state, due to its significant greater overlap with the wave packet, compared to when it is a bulk state. Thus approximately,
\begin{equation}
    \psi(\bm{r},t) \propto\sum_{n\in\text{edge}} \frac{c_n}{N} e^{-iE_n t} e^{(\ln\beta_y-\kappa_y)y+\ln(\beta_x)x} e^{ik_x^{(n)} x}.
\end{equation}

In the Hermitian limit, where $\gamma_{x,y}=0$ and $\beta_{x,y}=1$, the spacial dependent part in the kernel reduces to $e^{ik_x^{(n)}x-\kappa_yy}$, therefore the wave packet will stick to the bottom boundary until it reaches to the right edge, showing the typical chiral motion. As the non-Hermiticity gradually turns on, such that $\ln\beta_{x,y}>0$ but $\ln\beta_y-\kappa_y<0$, the wave packet remains attached to the bottom boundary as in the Hermitian limit [Fig.~\ref{f4}(a)]. However, its amplitude increases with the increasing of $x$, until it reaches the right boundary.

On the other hand, when the non-Hermiticity increases further such that $\ln\beta_y-\kappa_y>0$, the kernel of the wave packet is very similar to the topological trivial case in Eq.~\eqref{eq:topo_trivial}, which grows exponentially with both $x$ and $y$. Therefore, the wave packet will be detached from the bottom boundary and admit the bulk evolution, as shown in Fig.~\ref{f4}(b).

From the above analysis, it is clear that the competition between the NHSE and the intrinsic topological localization are equally important for determining the dynamical evolution of the wave packet. Explicitly, when $\ln\beta_y-\kappa_y<0$, the wave packet shows the boundary evolution; when $\ln\beta_y-\kappa_y>0$, the bulk evolution takes place.
In the following section, we analytically show this correspondence and derive the dynamical phase boundary line shown in Fig.~\ref{f4}(d).

\section{ IV. Derivation of the dynamical transition line} \label{sec:IV}
Based on the above analysis, the localization properties of the topological edge states play a crucial role in understanding the dynamical evolution of the wave packet in the non-Hermitian Chern insulators. Here, we propose an effective scheme to illuminate the competition between the intrinsic localization of edge states and non-Hermitian skin effects \cite{reply1}.

\begin{figure}[t]
\centering
\includegraphics[width=0.48\textwidth]{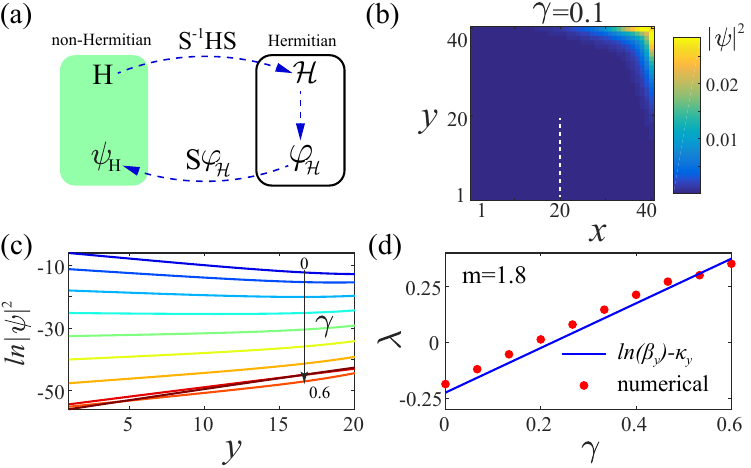}
	\caption{(a) A schematic illustrating the method for obtaining the topological edge states of $H$, using the transformed Hermitian Hamiltonian $\mathcal{H}=S^{-1}HS$ and its eigenfunction $\varphi_{\mathcal{H}}$ as the intermediate step. (b) A typical plot of the edge state with $\gamma=0.1$ and $m=1.8$. (c) The variation of the bottom edge states along the white dashed line marked in (b) for $\gamma\in[0,0.6]$. (d) The red dots are the slope of the edge states in (c), and the blue solid line is the analytical curve $\lambda=\ln\beta_y-\kappa_y$. Other parameters are $m=1.8$, $v=1$ and $\gamma_x=\gamma_y=\gamma$. }
	\label{f3}
\end{figure}

As illustrated in Fig. \ref{f3}(a), a similarity transformation $\mathcal{H}(\textbf{k})=S^{-1}H(\textbf{k})S$ transforms the non-Hermitian Hamiltonian $H(\textbf{k})$ into its Hermitian counterpart $\mathcal{H}(\textbf{k})$. For the Hermitian Hamiltonian, one is allowed to use the conventional method to obtain its edge states, obtaining $\mathcal{H} |\varphi_\mathcal{H}\rangle = E |\varphi_\mathcal{H}\rangle$.
Then, the topological edge states for the non-Hermitian Hamiltonian $H(\textbf{k})$ are determined by
\begin{align} \label{eq:reversed_transformation}
	\begin{split}
|\psi_{H}\rangle=S|\varphi_{\mathcal{H}}\rangle.	
\end{split}
\end{align}

For our case, the Hermitian counterpart of Eq.~\eqref{model} can be obtained as follows.
By using $S=\diag [1,\cdots,\beta_x^{x}\beta_y^{y},\cdots]$ with $\beta_{x,y}=e^{\gamma_{x,y}}$, Eq.~\eqref{model} can be rewritten as:
\begin{align}\label{hamiltonian2}
	\begin{split}
\mathcal{H}(\textbf{k})\!=\!v(\sin k_x\sigma_x\!+\!\sin k_y\sigma_y)\!+\!(m\!-\!\cos k_x\!-\!\cos k_y)\sigma_z.
	\end{split}
\end{align}
The above Hamiltonian $\mathcal{H}(\textbf{k})$ is exactly the transformed Hermitian Hamiltonian of Eq.~\eqref{model} by considering $k_{x,y}\rightarrow k_{x,y}-i\gamma_{x,y}$.
 Based on the GBZ theory, the above results demonstrate that the Hamiltonian of Eq.~\eqref{model} possesses the real energy spectrum under the open boundary condition, which is also the results of the pseudo-Hermiticity. Specifically, the Hermitian matrix $\mathcal{H}=S^{-1}HS$ leads to the equation $S^{-1}HS=S^\dagger H^\dagger(S^\dagger)^{-1}$, reflecting the pseudo-Hermitian property of $H$ with $H(SS^\dagger)=(SS^\dagger) H^\dagger$ \cite{PHsystem}.
 
For clarity, we still pay attention to the edge sates located at the bottom edge of the system.
The bottom-edge states of $\mathcal{H}(\textbf{k})$ can be written as
\begin{align}
	\begin{split}
\varphi_B(\bm{r})=\chi_y e^{ik_xx-\kappa_yy},
	\end{split}
\end{align}
where $\chi_y$ is the normalization two-component spinor. Substituting this trial wavefunction $\varphi_B(\bm{r})$ into the eigenvalue equation, we can obtain the intrinsic localization coefficient [see Appendix I for more details]
\begin{equation}\label{kappa}
\kappa_y=v- \sqrt{v^2+2m-4}.
\end{equation}
By the reversed transformation in Eq.~\eqref{eq:reversed_transformation}, the bottom topological edge states for the non-Hermitian cases $H(\textbf{k})$ read \cite{localization}:
\begin{align}\label{eq:psi_L}
	\begin{split}
\psi_B(\bm{r})\propto e^{(\ln{\beta_y}-\kappa_y)y}e^{(\ln\beta_x+ik_x)x},
	\end{split}
\end{align}

In Fig.~\ref{f3}(b), we present the numerical results for the probability density $|\psi|^2$ of the typical edge state of Eq.~\eqref{model}.
 To better show the localization property of the topological edge states, we focus on their amplitude modulation along the white dashed line in Fig.~\ref{f3}(b). As shown in Fig.~\ref{f3}(c), when $\gamma$ increases, $\ln |\psi|^2$ first linearly decreases with $y$ and then increases, indicating a transition of the topological edge states from exponential localization on the bottom to exponential localization on the top. The numerical data can be well fitted by a simple formula $\ln|\psi|=\lambda y+c$, with the localization coefficients $\lambda$ plotted as the red dots in Fig.~\ref{f3}(d). The analytical formula $\lambda=\ln\beta_y-\kappa_y$ from Eq.~\eqref{eq:psi_L} is also plotted in Fig.~\ref{f3}(d), which shows excellent agreement with the numerical data.

As shown in Figs. \ref{f3}(c) and (d) with $\ln \beta_y=\ln(e^{\gamma})=\gamma$, the topological edge states of $H(\textbf{k})$ exhibit the conventional localization properties where $(-\kappa_y +\gamma)<0$. Based on the discussion in the previous section, the evolution of the wave packet in this case should be along the boundary. On the other hand, when $(-\kappa_y +\gamma)>0$, the topologically protected non-Hermitian edge states lose their conventional localization properties, and the wave packet should spread into the bulk. This demonstrates a transition in their localization features, which distinguishes the regimes $(-\kappa_y +\gamma)<0$ and $(-\kappa_y +\gamma)>0$.
By setting $\lambda=\gamma-\kappa_y=0$ and considering Eq. (\ref{kappa}), the critical point of the localization transition for the bottom edge state satisfies
 \begin{align}\label{eq:dynamical_transition_line}
	\begin{split}
\gamma-\kappa_y=\gamma-v+ \sqrt{v^2+2m-4}=0.
	\end{split}
\end{align}
This result is consistent with the numerical findings in Fig. \ref{f3}(d), which corresponds to $\gamma_c \approx 0.225$ when $v=1$ and $m=1.8$.

In the $m\!-\!\gamma$ plane, Eq.~\eqref{eq:dynamical_transition_line} is plotted in Fig.~\ref{f4}(d), which shows excellent agreement with the separation between the boundary evolution ($P=1$) and the bulk evolution ($P=0$) of the wave packet, numerically calculated using the time-dependent Schr\"odinger equation. This demonstrates the correspondence between the static localization transition of edge states and the dynamical transition of the wave packet evolutions for the studied model, as discussed in the previous section.


\section{V. Discussion}
It is beneficial to clarify the distinctions between the conventional bulk-boundary correspondence and the static-dynamic correspondence revealed in this work. In general, the bulk-boundary correspondence is determined by the GBZ theory, which can be summarized as \cite{NHSE1,MGBZOS2}
\begin{equation}
\det[E-H_{\text{OBC}}]=\det[E-\mathcal{H}_{\text{OBC}}]\approx \det[E-\mathcal{H}_{\text{PBC}}].
\end{equation}
Here, OBC and PBC represent the open and periodical boundary conditions, respectively.
The main idea of GBZ theory is that by a similarity transformation $\mathcal{H}_{\text{OBC}}=S H_{\text{OBC}}S^{-1}$, the transformed Hamiltonian $\mathcal{H}$ satisfies the bulk-boundary correspondence with $\det[E-\mathcal{H}_{\text{OBC}}]\approx \det[E-\mathcal{H}_{\text{PBC}}]$. It ensures that $\mathcal{H}_{\text{PBC}}$ and $H_{\text{OBC}}$ share approximately the same eigenvalues as well as bulk band gaps. Thus, GBZ theory is very successful in obtaining the correct energy spectrum and topological transitions.

Our results examine how dynamical properties are influenced by the competition between NHSE and topological characteristics.
While GBZ theory is sufficient for characterizing NHSE-induced localization, it is inherently not designed to characterize the intrinsic localization of topological edge states.
As we have seen, the topological localization of topological edge states is also crucial for the system's dynamical evolution. Consequently, the topological phase boundary predicted by GBZ theory does not correspond well with the dynamical transition line of the system. This suggests that if one is interested in the system dynamics, one must not only apply GBZ theory but also consider the system's inherent properties, such as its intrinsic topological localization properties discussed in this work.

In our work, we mainly focus on a specific non-Hermitian model, showing a peculiar dynamical transition from edge to bulk even in the topological region. We emphasize that similar physical mechanisms should also be applicable in other topological models, where the intrinsic localizations of the topological edge states play an essential role   (Appendix. III).
Notably, the non-Hermitian Chern insulator has been experimentally realized \cite{NHCIexp1,NHCIexp2} very recently. The peculiar dynamical transition of the wave packet could potentially be realized in these systems. For the experimental detection, the real time dynamics of the wave packets as well as its propagation path can be detected in several platforms within the current experimental techniques \cite{observation1,observation2,observation3}.

\section{Acknowledgments}
We appreciate the valuable discussions with Hongfang Liu. This work was supported by the National Basic Research Program of China (Grants No. 2024YFA1409003), National Natural Science Foundation of China (Grants No. 12350401 and No. 12204044), and Shanghai Science and Technology Innovation Action Plan (Grant No. 24LZ1400800).

\section{Appendix I: derivation of the non-Hermitian edge states}

\begin{figure*}[t]
\centering
\includegraphics[width=0.70\textwidth]{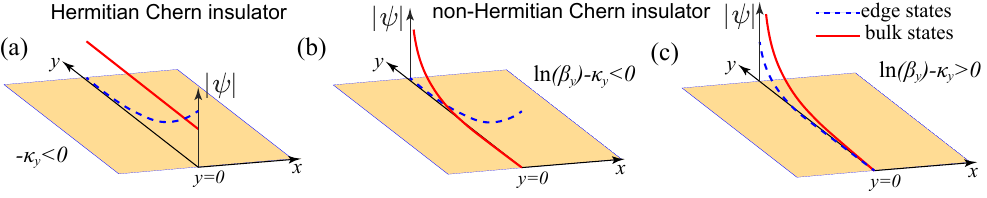}
	\caption{ Schematic plot of the sample. The dashed blue lines represent bottom edge states with $|\psi|\propto e^{[\ln (\beta_y)-\kappa_y]y}$. Here, $\ln (\beta_y)$ are induced by the NHSE. $\kappa_y$ is the intrinsic localization of edge states originating from the nontrivial Chern numbers.
(a) is the Hermitian cases with $\ln (\beta_y)=0$. (b) and (c) mark the typical edge and bulk states with $\ln (\beta_y)-\kappa_y<0$ and $\ln (\beta_y)-\kappa_y>0$, respectively. }
	\label{f2}
\end{figure*}

\textbf{Hermitian cases:} For $k \to 0$, $\sin (k)\sim k$ and $\cos (k)\sim (1-\frac{k^2}{2})$, Eq.~\eqref{hamiltonian2} can be written as
\begin{align}\label{Hermitianformula}
	\begin{split}
\mathcal{H}(\textbf{K})=&v k_x\sigma_x+v k_y\sigma_y+(m-2+\frac{1}{2} k_x^2+\frac{1}{2} k_y^2)\sigma_z.
	\end{split}
\end{align}
We focus on the bottom topological edge state localized along the $y$ direction [see Fig.~\ref{f2}(a)]. $\mathcal{H}(\textbf{k})$ can be rewritten as
$\mathcal{H}(\textbf{k})=h(k_x)+\delta h(k_y)$, with \cite{zongshu3}
\begin{align}
	\begin{split}
h(k_y)&=v k_y\sigma_y+(m-2+\frac{1}{2} k_y^2)\sigma_z,\\
\delta h(k_x)&=v k_x\sigma_x+\frac{1}{2} k_x^2\sigma_z.
	\end{split}
\end{align}
To calculate edge states localized in the $y$ direction, $\delta h(k_x)$ is not important and can be considered as the constant. By changing $k_y \to-i\partial_y$, we have
\begin{align}
	\begin{split}
h(-i\partial_y)&=-\frac{1}{2}\sigma_z \partial_y^2-iv\sigma_y\partial_y+(m-2)\sigma_z,
	\end{split}
\end{align}
Thus, the eigenequation for the bottom eigenstates writes as
\begin{align}
	\begin{split}
[-\frac{1}{2}\sigma_z \partial_y^2-iv\sigma_y\partial_y+(m-2)\sigma_z]\varphi_\mathcal{H}(y)=E\varphi_\mathcal{H}(y),
	\end{split}
\end{align}

Below, we mainly pay attention to the state $E\rightarrow0$, which possesses the largest localization length. By multiplying $\sigma_z$ from the left-hand side, one has
\begin{align}
	\begin{split}
[-\frac{1}{2} \partial_y^2-v\sigma_x\partial_y+(m-2)]\varphi_\mathcal{H}(y)=0.
	\end{split}
\end{align}
Assuming $\varphi_\mathcal{H}(y)=\chi_y e^{-\kappa_y y}$ and $y>0$ for the bottom edge states, the above equation can be rewritten as:
\begin{align}
	\begin{split}
[-\frac{1}{2} \kappa_y^2+ v\sigma_x\kappa_y+(m-2)]\chi_y=0.
	\end{split}
\end{align}
From the above equation, $\chi_y$ is the eigenvector of $\sigma_x$ , so that one must have $\sigma_x\chi_y=\eta\chi_y$ with $\eta=\pm1$. For $0<m<2$ and $v=1$, $\kappa_y>0$ dictates $\eta=1$, and we have
\begin{align}
	\begin{split}
-\frac{1}{2} \kappa_y^2+v\kappa_y+(m-2)=0.
	\end{split}
\end{align}
The localization coefficient $\kappa_x$ for the topologically protected edge is solved to be
\begin{align}
	\begin{split}
\kappa_y&=v- \sqrt{v^2+2m-4} .
	\end{split}
\end{align}
The minus sign before the square root is chosen,  due to the requirement of $\kappa_y\to 0$ when $m\to 2$. Since the bottom edge states are extended along the $x$ direction, the general form of $\varphi_{B}(\bm{r})$ can be written as
\begin{align}
	\begin{split}
\varphi_{B}(\bm{r})=\chi_ye^{ik_xx-\kappa_yy}.
	\end{split}
\end{align}

\textbf{Non-Hermitian counterpart:} Noticing $\beta_{x,y}=e^{\gamma_{x,y}}$,
the bottom edge states of the non-Hermitian Hamiltonian $H(\textbf{k})$ can be written as
\begin{align}
	\begin{split}
\psi_{B}(\bm{r})= S\varphi_{B}(\bm{r})=\chi_ye^{-\kappa_yy +\gamma_yy}e^{ik_xx +\gamma_xx}.
	\end{split}
\end{align}
The extra edges can be obtained in a similar manner. For illustration purpose, we schematically plot the variation of the edge state located at the bottom edge in Fig. \ref{f2}. It illuminates the competition between the intrinsic localization $\kappa_y$ and the NHSE $\ln (\beta_y)$.

Particularly, since the Hermitian Hamiltonian Eq.~\eqref{Hermitianformula} is obtained based on a similarity transformation of Eq. (\ref{model}) by $k_{x,y}\rightarrow k_{x,y}-i\gamma_{xy}$, they must have the same topological invariant based on the GBZ theory. Therefore, $\kappa_y=v- \sqrt{v^2+2m-4} =0$ is the topological phase transition condition for the non-Hermitian Chern insulator in Eq.~(\ref{model}). For $v=1$, the topological transition line is $m=2$, independent of $\gamma$.

Notably, the intrinsic localization lengths $\kappa_y=\kappa_y(E)$ of the edge state
are typically energy-dependent. When the largest intrinsic localization length of the edge state is comparable with $\ln \beta_y$, all the edge states lost their intrinsic localization properties. Consequently, dynamical evolution transitions from intrinsic-localization dominance to NHSE dominance.
Here, the largest localization length corresponds to the edge states at $E=0$, which are consistent with our deviations and the numerical results.

\section{Appendix II: Numerical approach to distinguish the evolution schemes}

\begin{figure*}[t]
\centering
\includegraphics[width=0.85\textwidth]{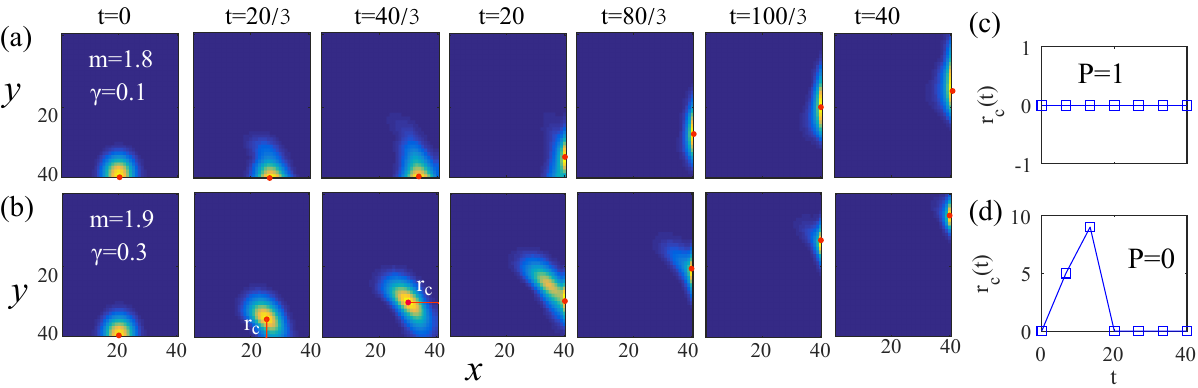}
	\caption{ (a) and (b) are the time evolutions of wave packet $|\psi(\bm{r}, t)|$ with boundary and bulk type evolutions, respectively. The red dots mark the wave packet centers with maximum values of $|\psi(\bm{r}, t)|$. (c) and (d) The variation of $r_c(t)$ versus $t$. The corresponding index $P$ are noted in the figures. The parameters are the same as those in the main text. The sample sizes are $N_x\times N_y=40\times 40$.}
	\label{R1}
\end{figure*}

The wave packet evolutions in Figs.~\ref{f4}(a) and Figs.~\ref{f4}(b) can be clearly distinguished by determining whether the wave packet center remains localized at the edge of the sample or not. Here, we clarify the calculation of $P$ in details.

Firstly, the evolution schemes are defined by identifing the evolution of the wave packet center. Its evolution satisfies the equation $\psi(\bm{r},t)=\langle\bm{r}|\psi(t)\rangle=\langle\bm{r}|e^{-iHt/\hbar}|\psi_{\text{in}}\rangle$. The wave packet center is located at $\bm{R}_c(t)$ where
\begin{equation}
\bm{R}_c(t)= \argmax_{x,y\in [1,N]}|\psi(\bm{r},t)|^2=[x_c,y_c].
\end{equation}

Secondly, one defines the
boundary coordinate for the set of sites located at the boundary, marked as $\bm{R}_b=[x_b,y_b]$. $\bm{R}_b$ can be easily identified based on the real-space coordinates in Fig. \ref{R1}(a). Here, $x_b$ and $y_b$ are a set of numbers. For example, for the lower boundary, one has $x_b\in[1,N_x]$ and $y_b=N_y$. $N_x$ and $N_y$ the sample sizes along the $x$ and $y$ direction respectively.

Thirdly, one defines the minimum distances between the wave packet center $\bm{R}_c(t)$ and the sample boundary $\bm{R}_b$
\begin{equation}
r_c(t)=\min|\bm{R}_c(t)-\bm{R}_b|=\min\{[x_c(t)-x_b]^2+[y_c(t)-y_b]^2\}.
\end{equation}
In Fig. \ref{R1}(b), a clear example of $r_c(t)$ can be identified for $t=\frac{40}{3}$. If the wave packet is located at the boundary, one has $r_c(t)=0$. Other wise, $r_c(t)>0$ demonstrate that the wave packets are mainly located in the bulk of the sample. The variation of $r_c(t)$ for the typical wave packet evolutions are presented in Figs. \ref{R1}(c) and (d). For simplicity, one adopts $R=\sum_t[r_c(t)]$ in our following calculations. For the edge (bulk) evolution, one has $R=0$ ($R\neq0$).

Fourthly, we define $(R==0)$ as a simple criteria,  which checks whether $R=0$ is true or not. For the edge evolution with $R=0$, (R==0) is true and gives the output $1$. Accordingly, for the bulk evolution with $R\neq0$, (R==0) is false and gives the output $0$.

Lastly, the index $P$ is defined as $P=[(R==0)]$. For the persistent edge evolution in Figs. \ref{R1}(a) and (c) with $R=0$, one has $P=1$. For the bulk evolution in Figs. \ref{R1}(b) and (d) with $R\neq0$, one has $P=0$.

In our calculations, the initial wave packet $\langle\bm{r}|\psi_{\text{in}}\rangle=e^{-\Delta_x(x-N/2)^2-\Delta_y(y-1)^2}$ is adopted with $\Delta_x=\Delta_y=1/40$. Its wave packet center is located at:
\begin{equation}
\bm{R}_c(t=0)=\argmax_{x,y\in [1,N]}|\psi_{\text{in}}(\bm{r},t=0)|^2=(\frac{N}{2},1),
\end{equation}
which is located at the bottom edge of the sample with $r_c(t=0)=0$.

By further counting the values of $r_c(t)$ for different times $t$, one is able to identify whether the wave packet evolves along the edge or spreads into bulk of the sample. Typically, if $r_c(t)$ matches the boundary cases for arbitrary $t$ with $r_c(t)=0$, one can conclude that $|\psi(t)\rangle=e^{-iHt}|\psi_{\text{in}}\rangle$ evolves along the edge of the sample [See Fig. \ref{R1}(c)]. Otherwise, $|\psi(t)\rangle$ spreads into the bulk [See Fig. \ref{R1}(d)]. These two cases give rise to: (i) $P=1$ for the persistent edge evolution; (ii) $P=0$ for the bulk evolution.

\begin{figure}[t]
\centering
\includegraphics[width=0.48\textwidth]{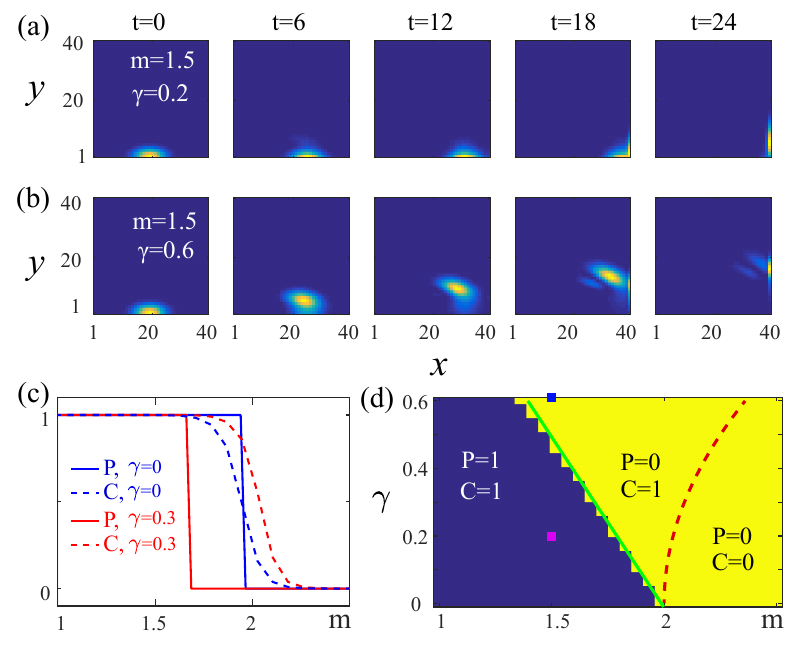}
	\caption{(a) and (b) are the time evolutions of wave packet $|\psi(\bm{r}, t)|$ with typical boundary and bulk evolutions, respectively. Note that they are both in the topological region, as marked in (d) with red and blue squares. (c) The variation of $P$ and $C$ for two representative $\gamma$'s, showing clear deviation for non-Hermitian case. (d) The plot of $P$ and $C$ versus $m$ and $\gamma$ with system size $N_x=N_y=30$. The red dashed line is the topological phase transition line based on the GBZ theory. The solid green line represents the dynamical transition line given by Eq.~\eqref{EQ31}. The parameters are $v=1$ , with others marked in the figure accordingly.}
	\label{A1}
\end{figure}

\section{Appendix III: An extra Chern insulator example}

To demonstrate the universality of the proposed scheme, we present the parallel results for an extra non-Hermitian Chern insulator model \cite{NHSE2}
\begin{align}
	\begin{split}
H(\textbf{k})&=(v\sin k_x+i\gamma)\sigma_x+(v\sin k_y+i\gamma)\sigma_y\\
&+(m-\cos k_x-\cos k_y)\sigma_z.
	\end{split}
\end{align}
The static properties of the Hamiltonian have been well studied \cite{NHSE2} with a topological phase transition line $m=2+\gamma^2$ [see the red dashed line in Fig. \ref{A1}(d)], which explicitly depends on $\gamma$. Based on similar calculations, the localization transition line for its edge states is
 \begin{align}\label{EQ31}
	\begin{split}
\gamma=2-m.
	\end{split}
\end{align}
It separates the dominance of intrinsic topological localization from the dominance of localization induced by non-Hermitian skin effects.

As shown in Figs.~\ref{A1}(d), the green solid line marks the localization transition boundary in Eq.~(\ref{EQ31}).
Notably, the analytical result fits the numerical result of dynamic transitions between $P=0$ and $P=1$ very well.
In addition, the dynamic transitions are obviously inconsistent with the topological transitions marked by the Chern numbers, as illustrated in Fig.~\ref{A1}(c) and Fig.~\ref{A1}(d).
The plot of the typical wave packet evolutions are presented in Figs.~\ref{A1}(a) and Fig.~\ref{A1}(b). Although they both possess the nontrivial Chern numbers, two distinct evolution schemes can be clearly identified. These results further demonstrate the validity of the peculiar dynamical transition in non-Hermitian Chern insulators.

\end{document}